\documentclass[twocolumn]{aastex631}

\usepackage{booktabs}
\usepackage{amsmath}
\usepackage{xcolor}

\graphicspath{{./}{figures/}}

\accepted{16 March 2023}

\submitjournal{ApJ} 

\shorttitle{Extreme Cooling and Star Formation at $z=1.2$ in SPT-CL J2215-3537}
\shortauthors{Calzadilla et al.}

\begin{document}

\title{SPT-CL J2215-3537: A Massive Starburst at the Center of the Most Distant Relaxed Galaxy Cluster}

\correspondingauthor{Michael Calzadilla}
\email{msc92@mit.edu}

\author[0000-0002-2238-2105]{Michael S. Calzadilla}
\affil{Kavli Institute for Astrophysics and Space Research, Massachusetts Institute of Technology Cambridge, MA 02139, USA}

\author[0000-0001-7665-5079]{Lindsey E. Bleem}
\affil{High Energy Physics Division, Argonne National Laboratory, 9700 S. Cass Avenue, Argonne, IL 60439, USA}
\affil{Kavli Institute for Cosmological Physics, University of Chicago, 5640 South Ellis Avenue, Chicago, IL 60637, USA}

\author[0000-0001-5226-8349]{Michael McDonald}
\affil{Kavli Institute for Astrophysics and Space Research, Massachusetts Institute of Technology Cambridge, MA 02139, USA}

\author[0000-0003-1370-5010]{Michael D. Gladders}
\affil{Kavli Institute for Cosmological Physics, University of Chicago, 5640 South Ellis Avenue, Chicago, IL 60637, USA}
\affil{Department of Astronomy and Astrophysics, University of Chicago, 5640 South Ellis Avenue, Chicago, IL 60637, USA}

\author[0000-0002-8031-1217]{Adam B. Mantz}
\affil{Kavli Institute for Particle Astrophysics and Cosmology, Stanford University, 452 Lomita Mall, Stanford, CA 94305, USA}

\author[0000-0003-0667-5941]{Steven W. Allen}
\affiliation{Kavli Institute for Particle Astrophysics and Cosmology, Stanford University, 452 Lomita Mall, Stanford, CA 94305, USA}

\author[0000-0003-1074-4807]{Matthew B. Bayliss}
\affil{Department of Physics, University of Cincinnati, Cincinnati, OH 45221, USA}

\author[0000-0003-2895-6218]{Anna-Christina Eilers}
\affil{Kavli Institute for Astrophysics and Space Research, Massachusetts Institute of Technology Cambridge, MA 02139, USA}

\author[0000-0003-4175-571X]{Benjamin Floyd}
\affil{Faculty of Physics and Astronomy, University of Missouri -- Kansas City, 5110 Rockhill Road, Kansas City, MO 64110, USA}

\author[0000-0001-7271-7340]{Julie Hlavacek-Larrondo}
\affil{Département de Physique, Université de Montréal, Succ. Centre-Ville, Montréal, Québec, H3C 3J7, Canada}

\author[0000-0002-3475-7648]{Gourav Khullar}
\affiliation{Department of Physics and Astronomy and PITT PACC, University of Pittsburgh, Pittsburgh, PA 15260, USA}

\author[0000-0001-6505-0293]{Keunho J. Kim}
\affil{Department of Physics, University of Cincinnati, Cincinnati, OH 45221, USA}

\author[0000-0003-3266-2001]{Guillaume Mahler}
\affil{Centre for Extragalactic Astronomy, Durham University, South Road, Durham DH1 3LE, UK}
\affil{Institute for Computational Cosmology, Durham University, South Road, Durham DH1 3LE, UK}

\author[0000-0002-7559-0864]{Keren Sharon}
\affiliation{Department of Astronomy, University of Michigan, 1085 S. University Ave, Ann Arbor, MI 48109, USA}

\author[0000-0003-3521-3631]{Taweewat Somboonpanyakul}
\affiliation{Kavli Institute for Particle Astrophysics and Cosmology, Stanford University, 452 Lomita Mall, Stanford, CA 94305, USA}

\author[0000-0003-0973-4900]{Brian Stalder}
\affil{Center for Astrophysics | Harvard \& Smithsonian, 60 Garden Street, Cambridge, MA 02138, USA}

\author[0000-0002-2718-9996]{Antony A. Stark}
\affil{Center for Astrophysics | Harvard \& Smithsonian, 60 Garden Street, Cambridge, MA 02138, USA}

\collaboration{18}{(SPT collaboration)}

\begin{abstract}
We present the discovery of the most distant, dynamically relaxed cool core cluster, SPT-CL J2215-3537 (SPT2215) and its central brightest cluster galaxy (BCG) at $z=1.16$. Using new X-ray observations, we demonstrate that SPT2215 harbors a strong cool core, with a central cooling time of 200 Myr (at 10 kpc) and a maximal intracluster medium cooling rate of $1900\pm400$ M$_{\odot}$ yr$^{-1}$. This prodigious cooling may be responsible for fueling extended, star-forming filaments observed in \textit{Hubble Space Telescope} imaging. Based on new spectrophotometric data, we detect bright [O \textsc{\lowercase{II}}] emission in the BCG, implying an unobscured star formation rate (SFR) of $320^{+230}_{-140}$ M$_{\odot}$ yr$^{-1}$. The detection of a weak radio source ($2.0\pm0.8$ mJy at 0.8 GHz) suggests ongoing feedback from an active galactic nucleus (AGN), though the implied jet power is less than half the cooling luminosity of the hot gas, consistent with cooling overpowering heating. The extreme cooling and SFR of SPT2215 is rare among known cool core clusters, and it is even more remarkable that we observe these at such high redshift, when most clusters are still dynamically disturbed. The high mass of this cluster, coupled with the fact that it is dynamically relaxed with a highly-isolated BCG, suggests that it is an exceptionally rare system that must have formed very rapidly in the early Universe. Combined with the high SFR, SPT2215 may be a high-$z$ analog of the Phoenix cluster, potentially providing insight into the limits of AGN feedback and star formation in the most massive galaxies.
\end{abstract}


\section{Introduction}\label{sec:intro}

Galaxy clusters are rich collections of hundreds to thousands of galaxies. However, most of the luminous mass in a cluster is found in a hot (T $\sim 10^7$ K), X-ray emitting phase that permeates the space between these galaxies. This vast X-ray emitting plasma is called the intracluster medium (ICM). As this ICM cools over time, it releases energy via Bremsstrahlung radiation and falls deeper into the gravitational potential well of the cluster. This cooling steepens the density profile of the cluster, which allows for more frequent interactions within the plasma and further decreases the temperature in the center, leading to what is referred to as a ``cool core" (CC) cluster \citep[e.g.][]{2010A&A...513A..37H}.

At the center of a CC cluster, there is usually one dominant, brightest cluster galaxy (BCG), onto which the cooling flow from the ICM is deposited. In the absence of any heat input, these BCGs are expected to vigorously form stars and contain large reservoirs of molecular gas \citep[see review by][]{1994ARA&A..32..277F}. Instead, observations of large samples of these systems show very little of either, with cooling suppressed, on average, by two orders of magnitude compared to theoretical predictions \citep[e.g.][]{1987MNRAS.224...75J, 1989AJ.....98.2018M, 1995MNRAS.276..947A, 1999MNRAS.306..857C, 2006ApJ...652..216R, 2008ApJ...681.1035O, 2015ApJ...805..177D, 2018ApJ...858...45M}. The cooling funneled onto the BCG also makes its way to its central active galactic nucleus (AGN), where accretion onto a supermassive black hole (SMBH) leads to massive outbursts in the form of relativistic outflows of plasma, which in turn deposit heat into the ICM. Such ``feedback'' in response to feeding establishes the central SMBH as a sort of thermostat that regulates the temperature and the amount of cool gas available for forming stars \citep[see reviews by][]{2012ARA&A..50..455F, 2007ARA&A..45..117M, 2012NJPh...14e5023M, 2020NatAs...4...10G, 2022PhR...973....1D}. In nearby clusters, this behavior seems tightly regulated \citep[e.g.][]{2015ApJ...805...35H}. We know much less about the behavior of feedback and cooling at higher redshifts, when clusters are still in the process of virializing and the availability of gas and the rate of cosmic star formation were much higher.

Studies of large samples of high-redshift clusters have been enabled by Sunyaev-Zeldovich (SZ) effect-based surveys \citep[e.g.][]{2010ApJ...722.1180V,2015ApJS..216...27B,2016A&A...594A..27P,2021ApJS..253....3H}, and more recently by more sensitive X-ray survey telescopes like eROSITA \citep{2022A&A...661A...2L}.
SZ surveys have discovered many more galaxy clusters in a mass-limited way, which allows us to study the balance between AGN feedback and ICM cooling over cosmic history \citep[e.g.][]{2013ApJ...774...23M}. 
Among the new lessons learned is that this feedback cycle has been in place for at least $\sim 10$ Gyr \citep{2015ApJ...805...35H,2022arXiv220713351R}. The fraction of cool cores in clusters has also remained constant with redshift \citep{2013ApJ...774...23M,2017ApJ...843...28M,2021ApJ...918...43R}. One might expect different behaviors at early times, as clusters are still rapidly forming and accreting subhaloes, and the average galaxy is more likely to be both star-forming and hosting an AGN \citep[e.g.][]{2013MNRAS.431.1638H,2022AJ....163..146S}.
One possible example of this is SpARCS104922.6+564032.5 (``SpARCS1049'', $z=1.7$), which hosts a starburst that is not centered on any galaxy., i.e. in the absence of AGN feedback \citep{2015ApJ...809..173W,2020ApJ...898L..50H}. Over time such a cooling flow may be quenched by the eventual alignment of the cluster potential between the BCG and AGN due to dynamical friction.

In this work we present the first detailed study of the massive (M$_{\rm 500c} = 7.32 \times 10^{14}$ M$_{\rm \odot}$) high-redshift ($z=1.16$) galaxy cluster SPT-CL J2215-3537 (hereafter SPT2215). This cluster was discovered in the SPTpol Extended Cluster Survey \citep[SPT-ECS;][]{2020ApJS..247...25B} and is among the most massive clusters known at $z>1$. Based on \textit{Chandra} imaging, \citet{2022MNRAS.510..131M} determined SPT2215 to be dynamically relaxed, and was considered as part of a larger sample of relaxed clusters in order to provide tighter constraints on cosmological parameters. This study focuses on some of the extraordinary properties of SPT2215 in more detail. 
In \autoref{sec:obs}, we describe multi-wavelength followup observations of this system and how these were reduced. In \autoref{sec:results}, we lay out the extreme qualities of this cluster, namely how isolated, relaxed, and star forming the central galaxy is. Finally, we discuss the importance of this cluster in a cosmological context in \autoref{sec:discussion} before summarizing in \autoref{sec:summary}. Throughout this paper, we assume a flat $\Lambda$CDM cosmology with $H_0 = 70$ km s$^{-1}$ Mpc$^{-1}$, $\Omega _m = 0.3$, and $\Omega _{\Lambda} = 0.7$. This yields a physical scale of 8.25 kpc arcsec$^{-1}$ at the redshift of the cluster. All measurement errors are 1$\sigma$ unless noted otherwise.

\section{Observations \& Data Reduction} \label{sec:obs}


\begin{figure*}[ht]
\includegraphics[width=\textwidth]{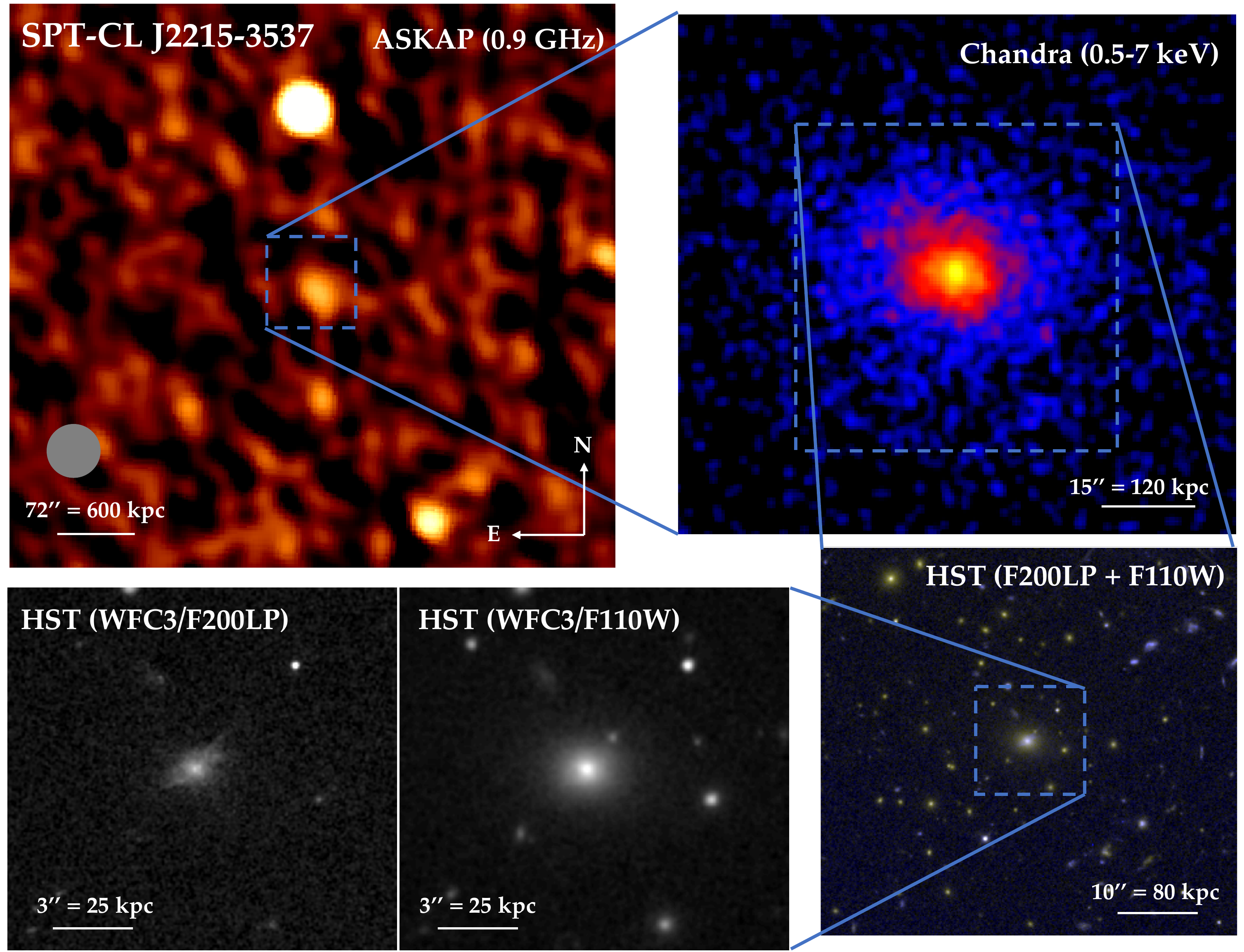}
\caption{Multiwavelength observations of SPT2215, progressively zooming in clockwise through the images. \textit{Top left}: ASKAP 887 MHz radio image cutout showing the presence of a radio source/AGN with an integrated flux of $2.0\pm0.8$ mJy at $0.8$ GHz. The ASKAP beam size is shown in gray. \textit{Top right}: $0.5 - 7$ keV \textit{Chandra} X-ray image showing a very relaxed ICM morphology and a strongly peaked central surface brightness profile. \textit{Bottom right}: 2-band \textit{HST} composite image of the cluster, with zoomed-in views of the central BCG in the \textit{bottom middle} (\textit{HST} WFC3/F110W) and \textit{bottom left} (\textit{HST} WFC3/F200LP) panels. The bluer F200LP observation shows the presence of extended, blue filaments that likely indicate strong star formation fueled by cooling out of the hot ICM.
\label{fig:multiwave_images}}
\end{figure*}

\subsection{X-ray (\textit{Chandra})}
\label{subsec:chandra}

\textit{Chandra} observations of SPT2215 were taken as part of a followup X-ray campaign of SPT SZ-selected clusters. The observations presented here were taken with the ACIS-I instrument, for a total of 72.26 ks (ObsIDs: 22653, 24614, 24615; PI: McDonald). These data were reduced and analyzed using the \textit{Chandra} Interactive Analysis of Observations {\tt (CIAO) v4.12.1} software with {\tt CALDB v4.9.2}, in a standard fashion similar to \citealt{2013ApJ...774...23M,2019ApJ...887L..17C,2021ApJ...918...43R,2022arXiv220713351R}.
The latest gain and charge-transfer inefficiency corrections were applied, with improved background screening for the \textsc{VFAINT} telemetry mode. Modeling of the global ICM properties was done as in \citet{2022arXiv220701624C}, and the thermodynamic profiles are shown and discussed in \autoref{subsec:coolcore}. 

To model the emission of the optically-thin, X-ray emitting plasma in this cluster, we use the \texttt{APEC/AtomDB XSPEC v3.0.9} thermal spectral model \citep{2001ApJ...556L..91S}, in addition to \texttt{PHABS} for photoelectric absorption \citep{1983ApJ...270..119M}. We adopt \citet{1989GeCoA..53..197A} abundances for consistency with previous literature, and Hydrogen column density $N_H = 1.07{\times}10^{20}$ cm$^{-2}$ from the Leiden-Argentine-Bonn survey \citep{2005A&A...440..775K}. 

\subsection{Optical Spectroscopy}
\label{subsec:spectroscopy}


As part of an ongoing follow-up campaign of SZ-selected clusters (e.g. \citealt{2016ApJS..227....3B,2019ApJ...870....7K,2022ApJ...934..177K}), the SPT collaboration has gathered optical spectroscopy of SPT2215 with the LDSS3 instrument on the 6.5m \textit{Magellan} telescopes in Chile. Multi-object spectra were obtained on 25 June 2019 using 1\farcs3 slits placed on the BCG and 16 other high-redshift member galaxy candidates.
Observations were made over 7 exposures totaling $2.3$h of LDSS3 spectroscopy with the \texttt{VPH-Red} grism ($1.175 {\rm \AA}$ pixel$^{-1}$ dispersion) and \texttt{open} filter, bracketed by flat and comparison \textit{HeNeAr} arc frames. The science and arc exposures were bias and flat-field corrected using \texttt{pyRAF/IRAF}\footnote{\url{https://iraf-community.github.io/pyraf.html}}, specifically with the \texttt{imred.ccdred} package. We also use the \textit{response} task from the \texttt{twodspec.longslit} to fit for the shape of the lamp spectrum in the dispersion direction before flat-fielding. In the resulting reduced science frames, we identify the BCG spectrum of SPT2215 and trace the slit using the \texttt{twodspec} package \textit{apall} to extract a one-dimensional (1D) spectrum with background subtraction. The same traced apertures were used to extract 1D spectra from the arc frames, which where then used to find a wavelength calibration solution with the \textit{identify} task. The \texttt{onedspec} \textit{refspec} and \textit{dispcor} tasks were then used to assign and apply the wavelength solution to the science frames. Each of the wavelength calibrated science frames were then median combined with cosmic ray rejection using \textit{scombine} to produce a final 1D spectrum of the BCG, with a wavelength range of $\lambda = [6800, 10500] ~ {\rm \AA}$, and dispersion of $2 {\rm \AA}$/pixel. 


\begin{figure*}[ht]
\centering
\includegraphics[width=\textwidth]{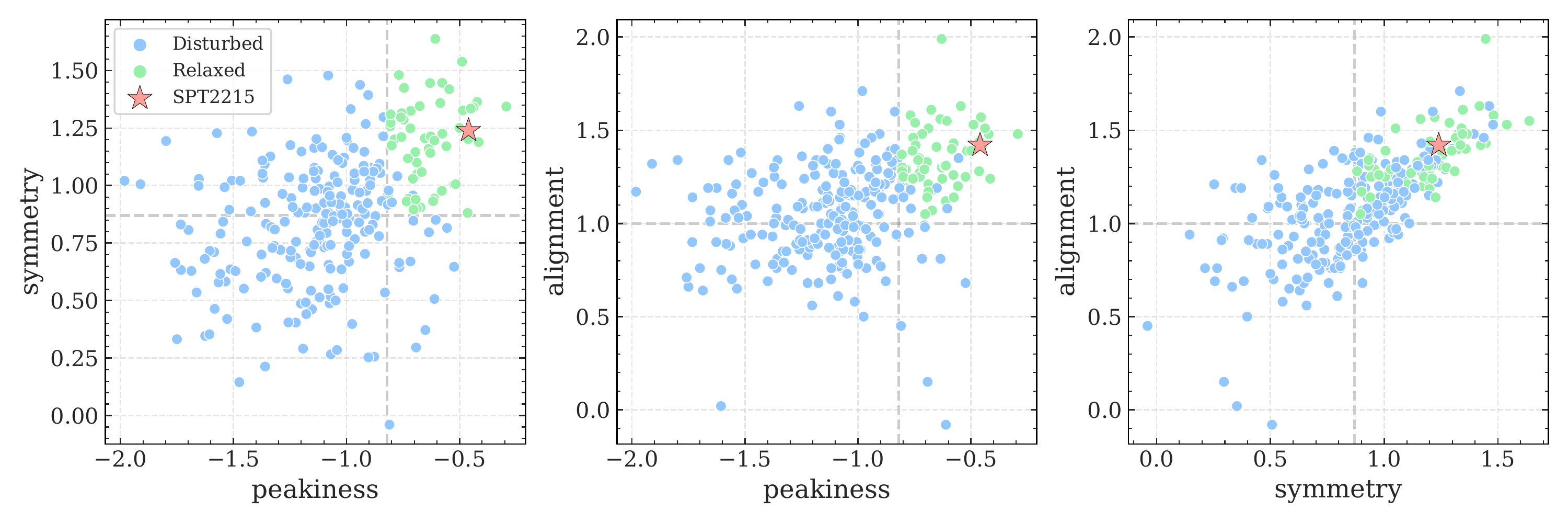}
\caption{SPA relaxedness criteria based on the symmetry (S), peakiness (P), and isophotal alignment (A) of ICM X-ray morphology for a sample of clusters \citep[adapted from][]{2015MNRAS.449..199M}. Combined three-dimensional cuts along each of these axes separate clusters into those that are dynamically relaxed (green circles) and unrelaxed (blue circles). SPT2215 is plotted as a red star, and can be seen to pass the above cuts distinguishing it as a relaxed cluster.
\label{fig:relaxed_SPA}}
\end{figure*}

\vspace{10mm}

\subsection{Optical/UV Imaging and Photometry}
\label{subsec:imaging}

\subsubsection{Hubble Space Telescope ACS/WFC3}

SPT 2215 was imaged as part of \textit{Hubble Space Telescope} Cycle 25, 26 \textit{HST}-SNAP program (GO-15307, GO-16017; PI: Gladders). Observations were obtained on UT 2017-10-06 with the Wide Field Camera 3 (WFC3) using the F110W filter (984s) and the F200LP filter (817s).
The F110W and F200LP filters sample the restframe emission from about $4170-6480 {\rm \AA}$ and $925-4600 {\rm \AA}$, respectively.
Special care had to be taken in the reduction of the F110W data to generate an appropriately flat fielded image reducing the effects of known `IR blobs,' which are small regions where detector sensitivity is lowered by $10-15$ \%. These features were modeled out by analyzing the stacked images of hundreds of similar exposure observations from the archive.
After this, the observations were processed in a standard way using STScI's \texttt{DrizzlePac} software package\footnote{\url{ https://drizzlepac.readthedocs.io/en/latest/}} with \texttt{AstroDrizzle  (v3.3.1)}.  Observations from both filters were combined and drizzled to the same pixel grid, with a resulting resolution of 0\farcs3 pixel$^{-1}$. All of the \textit{HST} data used in this paper can be found in MAST: \dataset[10.17909/pr0c-c679]{http://dx.doi.org/10.17909/pr0c-c679}.

\subsubsection{Magellan PISCO}
Optical imaging of SPT2215 were obtained for 300s in 1\farcs2 \ seeing on June 22, 2017 in the \textit{griz} bands using the Parallel Imager for Southern Cosmology Observations (PISCO; \citealt{2014SPIE.9147E..3YS})-installed on the 6.5 m \textit{Magellan} Clay Telescope. The image reduction process is detailed in \citet{2020ApJS..247...25B}. Sources were extracted using Source Extractor \citep{1996A&AS..117..393B}, star galaxy separation performed using the SG statistic \citep{2015ApJS..216...20B}, and photometric calibration performed using stellar locus regression (SLR) techniques \citep{2009AJ....138..110H}.
PISCO astrometry was tied to stars from the Dark Energy Survey public data release \citep{2018ApJS..239...18A}.

\subsubsection{Magellan FourStar}
To complement the optical imaging, 640s of \textit{J} and \textit{H} band near-infrared imaging with the \textit{Magellan/FourStar} \citep{2013PASP..125..654P} was obtained on Oct 02, 2017. As detailed in \citet{2020ApJS..247...25B}, the images were flat fielded with IRAF routines and astrometrically registered and relatively calibrated using the PHOTPIPE pipeline (e.g., \citealt{2007ApJ...666..674M}). Absolute photometric calibration was undertaken using SLR. Astrometry was tied to the Two Micron All Sky Survey (2MASS) catalog \citep{2006AJ....131.1163S}.

\subsubsection{Spitzer IRAC}
Spitzer/IRAC observations at 3.6 (\textit{I1}) and 4.5 $\mu$m  (\textit{I2}) of SPT2215 were obtained as part of a follow-up program to identify galaxy counterparts for high-redshift massive SZ cluster candidates (PID 11096, PI: Bleem). The cluster was observed for 360s on source time in both bands; the data were reduced and photometered following the methodology detailed in \citet{2009ApJ...701..428A}. Astrometry was also tied to the 2MASS catalog.

\section{Analysis \& Results}
\label{sec:results}

\subsection{The Most Distant Relaxed Cool Core Cluster}
\label{subsec:coolcore}

As mentioned before, SPT2215 was previously determined to be dynamically relaxed in \citet{2022MNRAS.510..131M}. This was done by characterizing the morphology of the X-ray emitting ICM, which looks largely spherically symmetric, as seen in the \textit{Chandra} image in the top right panel in \autoref{fig:multiwave_images}.
More quantitatively, using the ICM symmetry, peakiness, and isophotal alignment (SPA) criteria from \citet{2015MNRAS.449..199M} to determine the degree of relaxedness, these measurements yield
$S = 1.24 \pm 0.15$, 
$P = -0.46 \pm 0.04$,
$A =  1.42 \pm 0.16$ \citet{2022MNRAS.510..131M}.
In \autoref{fig:relaxed_SPA}, we compare the SPA measurements of SPT2215 to the sample of clusters from \citet{2015MNRAS.449..199M}, who identify thresholds that separate relaxed and disturbed clusters. We see that the SPA measurements for SPT2215 indeed lie firmly in the relaxed locus in this parameter space. 


The dynamical state of this cluster can be further validated by looking at the optical \textit{HST} images shown in \autoref{fig:multiwave_images} (bottom panels), which indicate that the BCG at the center of the frame is exceptionally bright, with an extended diffuse envelope of light. There are no other galaxies out to ${\sim} 200$ kpc in projected distance which appear nearly as bright or extended. We only consider this distance in order to minimize contamination from foreground or background galaxies not associated with the cluster. Within this radius, we measure background-subtracted F110W fluxes from ${\sim} 10$ kpc radius circular apertures centered on the brightest neighboring galaxies and compare to that of the BCG. The smallest magnitude gap, corresponding to the flux ratio between the BCG and the second brightest galaxy \citep[e.g.][]{2006ApJ...637L...9M,2007MNRAS.376..841V}, is $\Delta M_{1,2} \approx 2.2\pm0.1$. In the literature, systems with $R$ band gaps of $\Delta M_{1,2} \gtrsim 1.7$ are classified as fossil groups \citep[e.g.][]{2003MNRAS.343..627J}, and are thought to be old, undisturbed systems that have not experienced a significant recent merger \citep{2005ApJ...630L.109D}.
The large magnitude gap observed here suggests that SPT2215 has not recently experienced a significant merger, which is consistent with the exceptionally relaxed X-ray morphology.

\begin{figure*}[ht]
\centering
\includegraphics[width=0.8\textwidth]{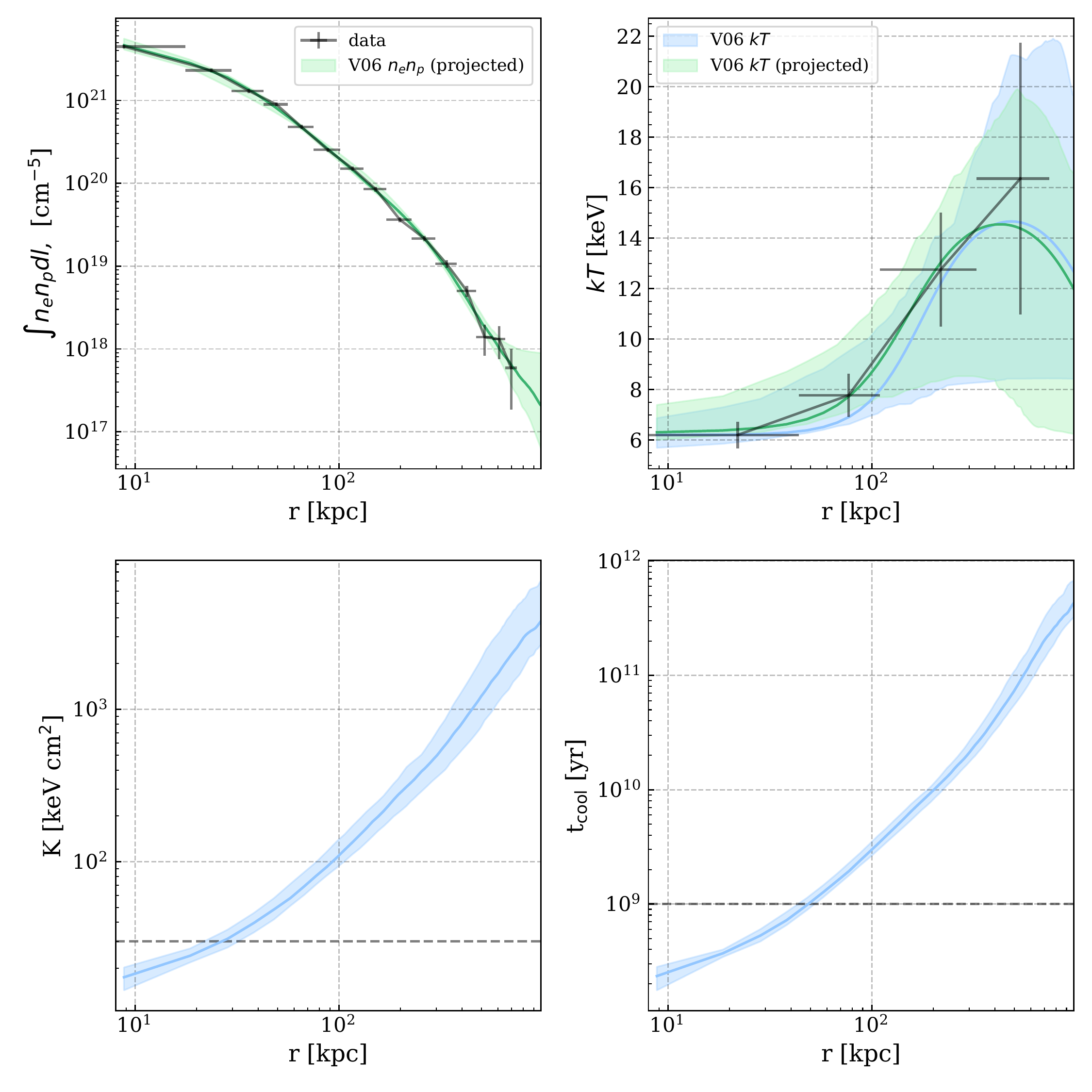}
\caption{ICM thermodynamic profiles for SPT2215. \textit{Top left}: emission measure (EM). \textit{Top right}: temperature ($k_BT$), in units of keV. \textit{Bottom left}: pseudo-entropy, in units of keV cm$^2$. \textit{Bottom right}: cooling time, in yrs. In the top panels, green shaded regions represent EM and temperature models from \citet{2006ApJ...640..691V} (eqns. 3 and 6 respectively) that have been projected along the line of sight and fit to the data which were sampled via the Monte Carlo method.
The corresponding unprojected (3D) models in all panels are shown as blue shaded regions. In the bottom panels, a horizontal dashed line marks the entropy ($K<30$ keV cm$^2$) and cooling time ($t_{\rm cool}<1$ Gyr) cool core thresholds below which we expect to see multiphase cooling, star formation, and AGN activity \citep[e.g.][]{2008ApJ...683L.107C}.
\label{fig:thermo_profiles}}
\end{figure*}

\begin{figure*}[ht] 
\includegraphics[width=\textwidth]{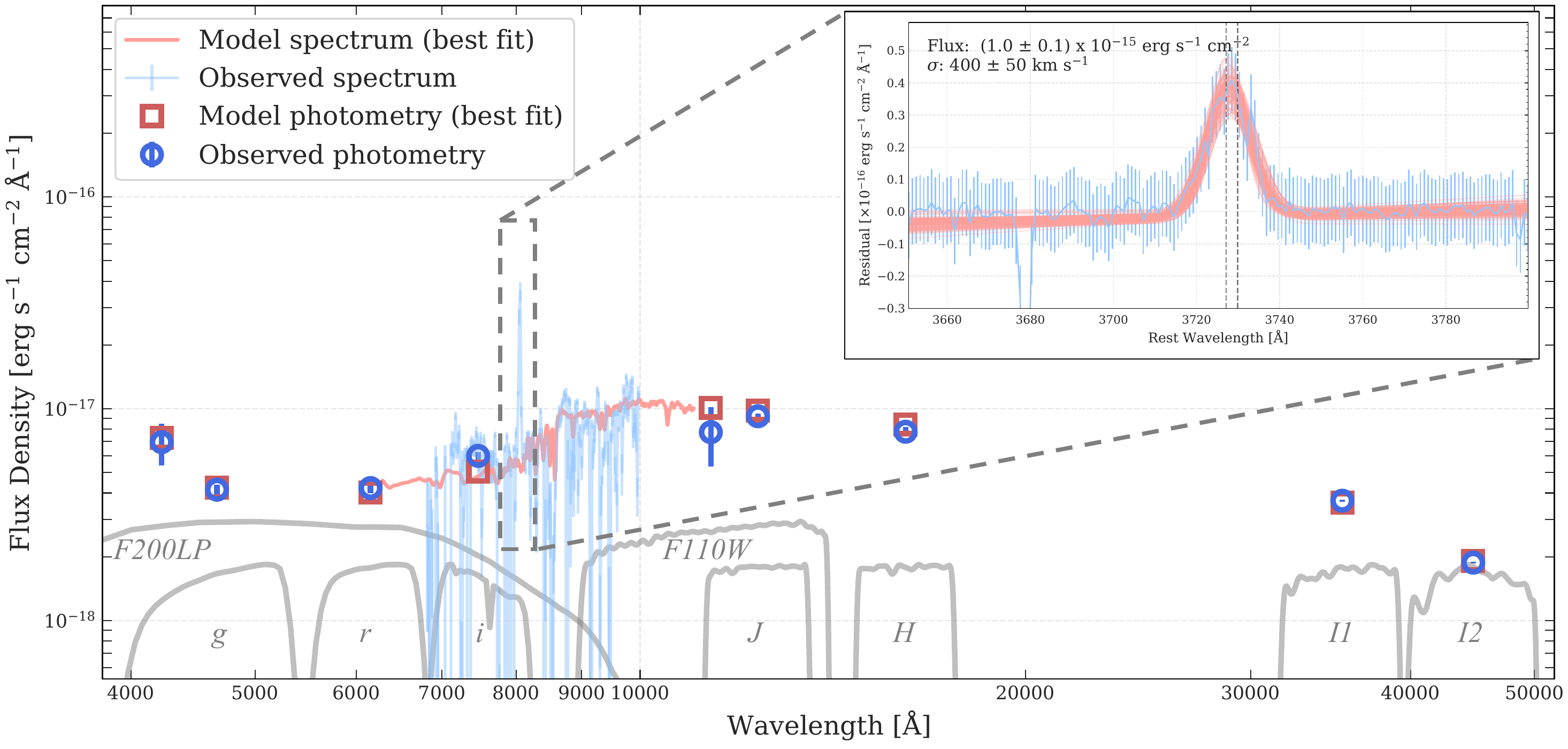}
\caption{Observed spectral energy distribution (SED) for the BCG in SPT2215, including the flux calibrated spectrum (light blue) and optical photometry (dark blue circles) from PISCO \textit{gri} bands, and F200LP and F110W \textit{HST} bands. Additional IR photometry in the \textit{FourStar} \textit{JH} bands, and \textit{Spitzer} \textit{I1} (3.6 $\mu$m) and \textit{I2} (4.5 $\mu$m) bands were also included in the SED fitting in order to better constrain the BCG stellar mass. The best fit model spectrum obtained with the \texttt{prospector} SED fitting software is shown in light red, with the best fit model photometry shown as dark red circles. The input spectrum was flux calibrated using the observed photometry to fit for the shape of the spectrum as a nuisance parameter, and binned by $30 {\rm \AA}$ here for clarity. In the inset, we show the unbinned, residual rest-frame spectrum after subtracting the best fit stellar continuum, zooming in on the spectral region surrounding strong [O \textsc{\lowercase{II}}] emission. We fit this emission to measure an obscured flux of ${\sim}1\times10^{-15}$ erg s$^{-1}$ cm$^{-2}$ which implies an unobscured SFR of $320^{+230}_{-140}$ M$_{\odot}$ yr$^{-1}$. In the rest frame, the doublet location
is consistent with the stellar continuum redshift (vertical dashed lines),
and has an instrumental broadening-corrected width of $400\pm50$ km s$^{-1}$, which could indicate the presence of a strong wind most likely powered by the massive ongoing starburst. 
\label{fig:spectrum}}
\end{figure*}

In addition to the morphological analyses above, we conduct a spectroscopic analysis of the X-ray data to determine thermodynamic properties of the ICM as a function of radius. We extract spectra from both coarse and fine annular bins centered on the BCG ($\alpha = 22^h15^m03\fs9306$, $\delta =  -35^{\circ}37\arcmin17\farcs885$). The coarse bins are used to fit a \texttt{PHABS$\ast$APEC} model with fixed metallicity ($Z=0.3Z_{\odot}$), in order to measure ICM temperature. We use a \citet{2006ApJ...640..691V} parametric model:

\begin{equation}
    T_{\rm 3D}(r) = T_0  \frac{\left(r/r_{\rm cool}\right)^{\alpha} + T_{\rm min}/T_0}{\left(r/r_{\rm cool}\right)^{\alpha} + 1}  \frac{\left(r/r_{\rm t}\right)^a}{[1 + \left(r/r_{\rm t}\right) ^b]^{c/b}}
\end{equation}
where $a$, $b$, and $c$ model the outer regions of the profile with a flexible broken power-law. The profile falls off at large radii at around $r_t$, while the inner cool core region is defined by $T_0$, $T_{\rm min}$, $r_{\rm cool}$, and $\alpha$. This 3D temperature profile is projected along the line of sight and fit to the extracted coarse temperature profile in projection. To quantify the uncertainty in our best fit model, we perform our fits over 100 Monte Carlo simulations that sample the data points within their errors, assuming Gaussianity.
As we only extract four spectroscopic temperature measurements, we freeze a number of parameters in our fits to be able to reliably constrain our temperature model. In all Monte Carlo iterations, we set $a=0$ and $\alpha=2$, while the parameters $b$, $c$, and $T_0$ are each randomly sampled in each iteration from the distribution of values found in \citet{2006ApJ...640..691V} (Table 3) and subsequently fixed. Additionally, $T_{\rm min}$ is fixed to the innermost randomly sampled temperature value for any given Monte Carlo iteration.
The best temperature fit is interpolated at the same radii as the fine annular binning scheme described before, for which we again extract spectra and fit them with the same model. For these finer spectral fits, we fix the interpolated temperature values and allow only the \texttt{APEC} model normalization to vary. These normalizations per unit area can be converted to an emission measure profile, to which we fit a projected \citet{2006ApJ...640..691V} emission measure (EM) model:

\begin{multline}
    n_{\rm e} n_{\rm p} (r) = n_0^2  \frac{\left(r/r_{\rm c}\right)^{-\alpha}}{\left[1+\left(r/r_{\rm c}\right)^2\right]^{3\beta - \alpha/2}}  \frac{1}{\left[1 + \left(r/r_{\rm s}\right)^{\gamma}\right]^{\epsilon/\gamma}} \\
    + \frac{n_{02}^2}{\left[1+\left(r/r_{\rm c2}\right)^2\right]^{3\beta_2}}
\end{multline}

which is a modified double-beta model with a cusp, rather than a flat core (defined by $n_0$, $r_{\rm c}$, $\alpha$, $\beta$), a steeper outer profile slope (defined by $r_{\rm s}$, $\gamma$, and $\epsilon$), and a cool core component (defined by $n_{\rm 02}$, $r_{\rm c2}$, and $\beta_2$). Here, $n_e$ and $n_p$ correspond to the electron and proton number densities, respectively. In our fits, $\gamma =3$ and $\epsilon =5$ remain fixed, and all other parameters are allowed to vary and initialized to typical parameters found in \citet{2006ApJ...640..691V} (Table 2). These fits to the temperature and EM profiles are performed over 100 Monte Carlo realizations in order to get uncertainty regions for the corresponding profiles by sampling data points to fit from their measured uncertainties (assuming Gaussianity).  

The resulting best fit EM and temperature profiles can be found in \autoref{fig:thermo_profiles}. After obtaining the density ($n_e$) profile from the (unprojected) EM profile, we combine these individual profiles to produce unprojected (3D) 
pseudo-entropy ($k_B T n_e ^{-2/3}$), 
and cooling time profiles ($t_{\rm cool} = \frac{(n_e + n_p) k_B T}{n_e n_p \Lambda(k_B T, Z)}$).
We used the cooling function $\Lambda(k_{\rm B}T, Z)$ from \citet{1993ApJS...88..253S} as parameterized by \citet{2001ApJ...546...63T}. 
Given the central temperature drop (a factor of ${\sim} 3\times$), and increase in density, we determine that this cluster does in fact contain a cool core. At a radius of 10 kpc, the central entropy and cooling time reach ${\sim}20$ keV cm$^2$ and $200$ Myr, respectively. There are also a number of diagnostic criteria that predict where multiphase cooling out of the ICM ensues. For instance, \citet{2008ApJ...683L.107C} were among the first to find a central entropy threshold in the cores of galaxy clusters of $K = 30$ keV cm$^2$, or equivalently, where $t_{\rm cool} \approx 1$ Gyr, below which strong H$\alpha$ and radio emission can be detected. For SPT2215, the cooling time profile drops below 1 Gyr at a radius of roughly $60$ kpc, and the entropy drops below 30 keV cm$^2$ at a radius of roughly $30$ kpc. Having met these criteria, we might expect a large amount of cooling multiphase gas to condense and cool out of the hot X-ray emitting phase. As an upper limit to the amount of material that can cool, we calculate a maximal cooling rate of $\dot{M}_{\rm cool} \equiv M_{\rm gas}(r<r_{\rm cool}) / t_{\rm cool} (r = r_{\rm cool}) = 1900\pm400$ M$_{\odot}$ yr$^{-1}$, by dividing the gas mass within some cooling radius by the cooling time at that cooling radius. Here, we choose $r_{\rm cool} = 116\pm15$ kpc as the radius where the cooling time is $t_{\rm cool} = 3$ Gyr to probe cooling closer to the core and for ease of comparison to literature $\dot{M}_{\rm cool}$ values. 
This high cooling rate puts SPT2215 in an extreme part of parameter space that only a handful of other clusters occupy \citep[e.g. Phoenix and RBS797;][]{2018ApJ...858...45M}.


\subsection{An Extremely High SFR}
\label{subsec:sfr}

Given the extreme cooling rate implied by the \textit{Chandra} X-ray data described in \autoref{subsec:coolcore}, we may also expect this condensation to fuel a large amount of star formation. Zooming in on the central BCG, we see in the bottom left panel of \autoref{fig:multiwave_images} a network of filaments in the F200LP filter, which probes restframe emission blueward of ${\sim} 4300 {\rm \AA}$. These filaments reach a maximum projected extent of ${\sim} 20$ kpc, though deeper observations could reveal more emission further out. Filamentary nebulae like these typically signal regions of ionization by young stars, and can be seen in many other strong cooling clusters \citep[e.g.][]{2010ApJ...721.1262M,2022arXiv220701624C}.
This observation utilizes a very broad F200LP filter 
that includes a large amount of rest frame UV continuum, so we only use it here for an approximate estimate of the UV-derived SFR.
By measuring the flux from a circular aperture ($2\farcs5 \approx 20$ kpc in radius) centered on the BCG, we measure a UV luminosity of $(3.7\pm0.3) \times 10^{29}$ erg s$^{-1}$ Hz$^{-1}$. This luminosity can be converted to a SFR using the \citet{2002MNRAS.332..283R} relation, which is calibrated to account for any intrinsic extinction when it is not possible to measure it directly. With this relation, we calculate a SFR of ${\sim} 240\pm20$ M$_{\odot}$ yr$^{-1}$ .

The high resolution but shallow \textit{HST} imaging data in \autoref{fig:multiwave_images} are complemented by ground-based imaging and spectroscopy using the \textit{Magellan} 6.5m telescopes. 
In particular, our LDSS3 spectroscopic observations for this cluster are especially helpful in securing an independent optical SFR estimate, as well as a precise redshift. 
The wavelength-calibrated spectrum obtained from the reduction steps outlined in \autoref{subsec:spectroscopy} was fit in combination with optical \textit{gri} photometry from PISCO and the F110W and F200LP bands from \textit{HST}, as well as IR photometry using the \textit{JH} bands from \textit{FourStar} and \textit{I1}
and \textit{I2} bands from \textit{Spitzer}. The IR bands are especially helpful in constraining the galaxy mass since most of the redder light from the bulk of the stars is redshifted out of the optical bands. This spectrophotometry was fit using the SED fitting code \texttt{prospector} \citep{2021ApJS..254...22J}, primarily in order to correct the shape of the spectrum, which we show in \autoref{fig:spectrum}.
\texttt{Prospector} uses Markov Chain Monte Carlo (MCMC)-based stellar population synthesis (SPS) based on the Python-FSPS framework \citep{2009ApJ...699..486C,2014zndo.....12157F}, as well as the MILES stellar spectral library \citep{2011A&A...532A..95F} and MIST isochrones \citep{2016ApJ...823..102C}. 
Our SED modeling of the stellar continuum is described by 
a delayed-tau parametric star formation history of the form ${\rm SFR}(t,\tau) \propto \left(t/\tau\right) e^{-t/\tau}$, as in \citet{2022ApJ...934..177K}, with additional free parameters to capture a burst of recent star formation as well as dust attenuation following \citet[][eqns. 1-3]{2013ApJ...775L..16K}, with a variable dust index of $\delta$.
A description of the priors for each of these parameters may be found in \autoref{tab:priors}. To fit only the stellar continuum, we mask parts of the spectrum associated with potential emission lines from [O \textsc{\lowercase{II}}], [O \textsc{\lowercase{III}}] and the Balmer series for Hydrogen. From these SED fits, we get a best fit redshift of $z=1.1598 \pm 0.0005$. We also measure a BCG remnant stellar mass of $\log(M_{\rm BCG}/M_{\odot}) = (11.77\pm0.04)$, with stellar metallicity $\log(Z/Z_{\odot}) = -0.20\pm0.14$, an age of $t = 4.5\pm0.5$ Gyr, and a star formation e-folding time of $\tau = 1.56\pm0.28$ Gyr. The starburst is estimated to have formed $11\pm9$\% of the remnant stellar mass, starting at $64\pm16$\% of the BCG's age. For the dust extinction associated with the old stellar continuum, we measure $A_{V,s} = 0.68\pm0.15$, and an additional gas or young stellar extinction of $A_{V,g} = 1.37\pm0.41$.

The best fit stellar continuum (seen in \autoref{fig:spectrum}) is subtracted from this flux calibrated spectrum, and in the residual spectrum we look for the presence of [O \textsc{\lowercase{II}}] which indicates star formation \citep[e.g.][]{1998ARA&A..36..189K,2004AJ....127.2002K}. We shift the residual observed spectrum to the rest-frame and fit a double Gaussian to this [O \textsc{\lowercase{II}}]$\lambda\lambda3726,2739$ doublet, tying the velocity dispersions and wavelengths of the two lines, as well as a straight line to approximate any remaining  residual continuum. Our fit in \autoref{fig:spectrum} (inset) yields an [O \textsc{\lowercase{II}}] flux of ($1.0 \pm 0.1) \times 10^{-15}$ erg s$^{-1}$ cm$^{-2}$, or a luminosity of $2.6 \times 10^{43}$ erg s$^{-1}$. 
To appropriately calculate an intrinsic (dereddened) flux, we use the extinction measurement for young stars ($A_{V,g}$) cited above, which results in a color excess of $E(B-V) = 0.34\pm0.10$, assuming $R_V=4.05$. 
We thus estimate an intrinsic [O \textsc{\lowercase{II}}] flux of ($6.4^{+4.6}_{-2.7}) \times 10^{-15}$ erg s$^{-1}$ cm$^{-2}$, or a luminosity of $(4.8^{+3.5}_{-2.0}) \times 10^{43}$ erg s$^{-1}$, and use the SFR-$L_{\rm [O \textsc{\lowercase{II}}]}$ relation from \citet{2004AJ....127.2002K} to calculate a SFR of $320^{+230}_{-140}$ M$_{\odot}$ yr$^{-1}$. This SFR is in good agreement with the $600 \pm 110$ M$_{\odot}$ yr$^{-1}$ predicted from the empirically-calibrated relation between unobscured SFR and obscured [O \textsc{\lowercase{II}}] luminosity from  \citet{2002MNRAS.332..283R}, which can be used when a direct measurement of the extinction is not possible. For robustness, we also tried our SED fitting with the dust attenuation curve of \citet{2000ApJ...533..682C}, and found a similar extinction and SFR estimate.
As a final check, these optically-derived SFR estimates are also consistent with the independent UV SFR estimate inferred from the shallow \textit{HST} data mentioned above. 
Such a high level of star formation makes SPT2215 stand out among most CC clusters, with only a handful of others exceeding $100$ M$_{\odot}$ yr$^{-1}$.

Another noteworthy feature of the [O \textsc{\lowercase{II}}] detection is the broad line width. With an instrumental broadening-corrected rest frame width of $\sigma = 400\pm50$ km s$^{-1}$ -- compared to a typical dispersion of roughly $150\pm100$ km s$^{-1}$ \citep[e.g.][]{2016MNRAS.460.1758H,2018ApJ...854..167G} -- the [O \textsc{\lowercase{II}}] emission could be a potential outflow/wind. 
The total 2D spectrum does not show much structure in the emission in the spatial direction, so a rotation scenario seems unlikely. For comparison, the spatially-integrated spectrum of the Phoenix cluster has a linewidth of $\sigma \sim 350$ km s$^{-1}$ \citep{2012Natur.488..349M}. A spatially-resolved spectroscopic study of SPT2215 would allow us to understand the origin of this broad, spatially-extended emission feature. 




\section{Discussion} 
\label{sec:discussion}

The fact that SPT2215 is a strong cool core at such a high redshift gives us a new, unique window into AGN feeding and feedback. In order to investigate the ``feedback’’ side, we look for the presence of a radio source. We find a faint radio source (see \autoref{fig:multiwave_images}, top left panel) with an integrated flux density of $2.0\pm0.8$ mJy at 0.8 GHz detected with the ASKAP radio array \citep{2008ExA....22..151J}. This is below the SUMSS (6 mJy beam$^{-1}$ limit) detection limit, and as such it is not detected in this survey. After k-correction using a spectral index\footnote{where flux density $S_\nu$ scales with frequency as $S_\nu \propto \nu^\alpha$} of $\alpha = -0.7$, the 0.8 GHz ASKAP flux corresponds to a 1.4 GHz radio luminosity of $9.3 \times 10^{40}$ erg s$^{-1}$. Though our \textit{Chandra} X-ray data are not deep enough to detect potential cavities and directly calculate an associated cavity power ($P_{\rm cav} $), we can use the 1.4 GHz radio luminosity and the scaling relation from \citet{2010ApJ...720.1066C} to estimate $\log$$(P_{\rm cav}/10^{42}$ erg s$^{-1}$) $= 2.6 \pm 0.3$ (stat.) $\pm 0.8$ (int. scatter).
From our X-ray analysis, we also estimate a cooling luminosity of 
$L_{\rm cool} = (3.7\pm0.1) \times 10^{45}$ erg s$^{-1}$, 
which is the bolometric ($0.01 - 100$ keV) X-ray luminosity measured within a radius of $116$ kpc, where the $t_{\rm cool}$ profile reaches 3 Gyr.
These measurements yield a ratio of $\log (P_{\rm cav} /  L_{\rm cool}) = -1.0\pm0.8$ (where the uncertainty is dominated by the scatter in the $P_{\rm cav}$-$L_{\rm 1.4 ~ GHz}$ relation), which suggests that the ICM cooling is overwhelming AGN feedback in this system by an order of magnitude. This measured $P_{\rm cav} /  L_{\rm cool}$ ratio is suggestive of a strong imbalance between heating and cooling, but is ultimately consistent with the large amount of scatter found in $P_{\rm cav} /  L_{\rm cool}$ as a function of redshift \citep{2022arXiv220713351R}.

With a massive cool core and a $P_{\rm cav} /  L_{\rm cool} \approx 0.1$, it may be no surprise that we measure such a high SFR. Perhaps more noteworthy is the fact that, compared to its maximal cooling rate of $\dot{\rm M}_{\rm cool} \sim 1900$ M$_{\odot}$ yr$^{-1}$, we calculate a cooling conversion efficiency of $\epsilon_{\rm cool} \equiv$ SFR/$\dot{\rm M}_{\rm cool} = 17\pm10$ \%. 
It may be argued that the radius where we measure $L_{\rm cool}$ and $\dot{\rm M}_{\rm cool}$ is somewhat arbitrary, so for comparison, we calculate these at the radius of ${\sim}20$ kpc, where we actually observe the cooling filaments (\autoref{fig:multiwave_images}, lower left panel). At this smaller radius, $t_{\rm cool, 20} \approx 0.3$ Gyr,  $\dot{\rm M}_{\rm cool, 20} = 610\pm220$ M$_{\odot}$ yr$^{-1}$, $L_{\rm cool, 20} = (8.4\pm0.5) \times 10^{44}$ erg s$^{-1}$, and the resulting cooling conversion efficiency is much higher at $\epsilon_{\rm cool, 20} = 52\pm36$ \%.
In most low-redshift systems, cooling is suppressed by about two orders of magnitude on average, with $\epsilon_{\rm cool} \sim 1$\%. However, \citet{2022arXiv220701624C} find that $\epsilon_{\rm cool}$ scales directly with $\dot{\rm M}_{\rm cool} $ \citep[see also][]{2017ApJ...846..103F,2018ApJ...858...45M}. The high value of $\epsilon_{\rm cool}$ measured in SPT2215 is consistent with this trend towards increasingly efficient cooling in the strongest cool cores, though the reason for this trend is still not understood \citep{2022arXiv220701624C}.

The simplest explanation for the high star formation rate in SPT2215 is that cooling is exceeding heating in this system. This is supported by the observation that the cooling luminosity is a factor of two higher than the jet power, which ought to lead to a large residual cooling flow. However, the jet power here is being constrained via the radio luminosity, which can vary by factors of several on $\sim$10 year timescales \citep[e.g.,][]{2014MNRAS.442.2048D}, which makes it difficult to reliably connect it to the time-averaged jet power. With deeper observations, we could detect or put limits on the presence of X-ray cavities, which would provide an estimate of the long term jet power output, providing a better basis for comparison to the cooling luminosity.

Regardless of what the true $P_{\rm cav} / L_{\rm cool} $ ratio is, a significant power imbalance in one system is unlikely to explain the overall trend towards increasingly efficient cooling in the most massive cool cores such as SPT2215. An alternative explanation relates to the idea that cool cores are in a stable cycle between cooling and feedback. If accretion of cold gas onto a supermassive black hole can trigger jets which, in turn, can lift low-entropy gas to larger radii in the cluster, then cooling will be suppressed for roughly a rise and freefall time, before the cycle starts anew \citep[e.g.,][]{2020MNRAS.495..594P}. If this system is like a harmonic oscillator between cooling cycles, then the frequency of oscillation will depend on the depth of the potential. That is, the low-entropy gas will return to the cluster center much faster in a cluster with a deeper central potential than in a cluster with a shallow central potential. If this freefall timescale is shorter than the timescale for consuming the reservoir of cold, dense gas via star formation, then that reservoir will constantly be replenished, leading to higher average star formation rates compared to systems where star formation can cycle between on and off. This is a hypothesis that we can, and intend to, test via the combination of state-of-the-art simulations and data for a large variety of cool core clusters \citep[e.g.][]{2022arXiv220701624C}.

\begin{figure}[t]
\includegraphics[width=\columnwidth]{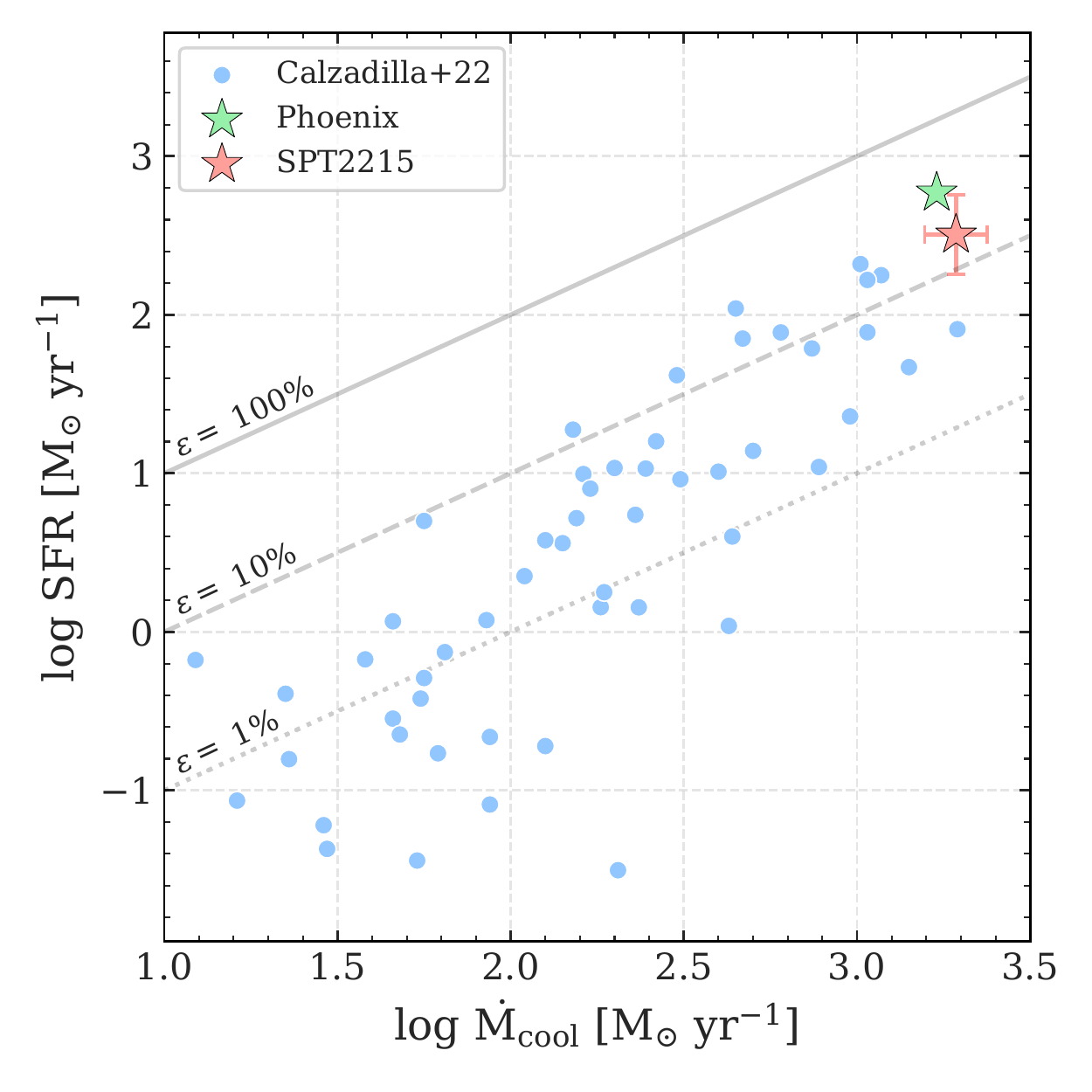}
\caption{SFR vs maximal cooling rate ($\dot{\rm M}_{\rm cool}$, at a radius where $t_{\rm cool} = 3$ Gyr) data from \citet{2022arXiv220701624C}, with SPT2215 overlaid, in addition to the Phoenix cluster. Lines of constant cooling efficiency ($\epsilon \equiv$ SFR/$\dot{\rm M}_{\rm cool}$) indicate that over $10$\% of SPT2215's maximum possible ICM cooling contributes into forming stars.
\label{fig:sfr_vs_mdot}}
\end{figure}

SPT2215 is similar to many of the most well-studied, extreme cool core clusters at low redshifts, but it is rare in that we observe it at such high redshift. As a point of contrast, it was recently argued that the even higher redshift ($z=1.7$) galaxy cluster SpARCS1049 was rapidly forming stars (SFR ${\sim} 860$ M$_{\odot}$ yr$^{-1}$) from a cooling flow, facilitated by a recent dynamical encounter that physically separated the low-entropy gas from the central galaxy, allowing for cooling in the absence of local feedback \citep{2015ApJ...809..173W,2020ApJ...898L..50H}. Over time, SpARCSS1049 may be expected to relax, leading to suppression of the cooling flow once it is aligned with and directed towards the BCG and central SMBH. In this sense, SPT2215 may be a potential intermediate step between a system like SpARCS1049 and other relaxed low-redshift systems, as the ICM cooling is centered on a BCG and there are signs of weak radio emission, possibly from a very recently ``activated’’ AGN. 

On the other hand, the high level of relaxedness in SPT2215 implies that, if a dynamical interaction triggered some initial cooling episode, it likely happened long ago. More realistically, SPT2215 and SpARCS1049 may represent two distinct avenues to rapid cooling of the ICM, either via a strong dynamical encounter that physically separates low-entropy gas from the source of feedback, or via a potential so deep that the freefall time of uplifted gas is shorter than the consumption time of the cold reservoir.
Instead, SPT2215 seems to be a high-$z$ analog of the Phoenix cluster, not only in how intensely star forming it is, but also in how massive and relaxed it is. Our SED fitting results in \autoref{subsec:sfr}  suggest the BCG itself is exceptionally massive, with a surviving stellar mass of $M_{\rm BCG} = (5.9\pm0.6) \times 10^{11}$ M$_{\odot}$. This inferred mass is consistent with that predicted by the scaling relations between stellar mass and the combination of \textit{Spitzer} magnitudes and $g-r$ colors found in \citet[][eqns. 6-7]{2010RAA....10..329Z}. Such an extraordinary mass at this redshift implies that SPT2215 may have had a head start in its accretion history, allowing it to grow extremely quickly. This is a trait that SPT2215 shares with the Phoenix cluster, which is also exceptionally massive and shows no sign of a recent merger, indicating rapid, early growth. Why such a rapid growth followed by a period of relaxedness ought to lead to over-cooling of the core remains an open question, but it seems unlikely to be a coincidence.

\section{Summary} 
\label{sec:summary}

SPT2215 is a unique cluster in that it is the most distant yet found of a remarkable set of extreme cool cores.
Using multiwavelength imaging and spectral observations, we have demonstrated that this system consists of a strongly cooling X-ray core (at a radius of 10 kpc, the central entropy and cooling time reach ${\sim}20$ keV cm$^2$ and $200$ Myr, respectively, and the maximal ICM cooling rate at a radius where the cooling time reaches 3 Gyr is $\dot{\rm M}_{\rm cool} \sim 1900$ M$_{\odot}$ yr$^{-1}$). This cool core is at the center of a dynamically relaxed cluster, which appears to be fueling a massive starburst ($320^{+230}_{-140}$ M$_{\odot}$ yr$^{-1}$) in a highly-evolved central giant elliptical galaxy. Further X-ray observations (Mantz et al. in prep.) will allow more precise thermodynamic profile modeling, measurement of gas fraction, and further constraints on cosmological models. Our shallow \textit{HST} data already reveal hints of complex star forming filaments, which will benefit from deeper observations with \textit{HST} or perhaps even \textit{JWST/NIRSpec} IFU in order to more precisely measure SFRs and reddening, and further understand the underlying process of cluster star formation in a previously untapped redshift regime.

\bigskip 
Based on observations made with the NASA/ESA {\it Hubble Space Telescope}, obtained at the Space Telescope Science  Institute, which is operated by the Association of Universities for Research in Astronomy, Inc., under NASA contract NAS 5-26555. These observations are associated with program numbers 15307, 15661, and 16001. The authors acknowledge support from programs HST-GO-15307.008-A, HST-GO-15661.001, and HST-GO-16001.002, provided through grants from the STScI under NASA contract NAS5-26555.
The authors also acknowledge additional financial support for this work, provided by the National Aeronautics and Space Administration through \textit{Chandra} Award Number GO0-21124A issued by the \textit{Chandra} X-ray Center, which is operated by the Smithsonian Astrophysical Observatory for and on behalf of the National Aeronautics Space Administration under contract NAS8-03060. 
The South Pole Telescope program is supported by the National Science Foundation (NSF) through grants PLR-1248097 and OPP-1852617. Partial support is also provided by the NSF Physics Frontier Center grant PHY-1125897 to the Kavli Institute of Cosmological Physics at the University of Chicago, the Kavli Foundation, and the Gordon and Betty Moore Foundation through grant GBMF\#947 to the University of Chicago. Argonne National Laboratory’s work was supported by the U.S. Department of Energy, Office of Science, Office of High Energy Physics, under contract DE-AC02-06CH11357.
This paper used data gathered with the 6.5 m \textit{Magellan} Telescopes located at Las Campanas Observatory, Chile. We thank the staff of Las Campanas for their dedicated service, which has made these observations possible. 
The scientific results reported in this article are based on observations made by the \textit{Chandra} X-ray Observatory, and this research has made use of software provided by the \textit{Chandra} X-ray Center (CXC) in the application package, CIAO.
MSC acknowledges support from the NASA Headquarters under the Future Investigators in NASA Earth and Space Science and Technology (FINESST) award 20-Astro20-0037. 
AAS acknowledges support from NSF award 2109035. 
GM acknowledges funding from the European Union's Horizon 2020 research and innovation program under the Marie Skłodowska-Curie grant agreement No MARACHAS - DLV-896778.


\facilities{CXO, \textit{HST}. NSF/US Department of Energy 10m South Pole Telescope (SPTpol). Magellan 6.5m Telescopes 
(Clay/LDSS3C, Clay/PISCO, Baade/FourStar)}

\software{astropy \citep{2013A&A...558A..33A}, 
numpy \citep{2020Natur.585..357H}, 
scipy \citep{2020NatMe..17..261V}, 
pandas \citep{2022zndo...3509134R}, 
CIAO \citep{2006SPIE.6270E..1VF}, 
XSPEC \citep{1996ASPC..101...17A}. 
Prospector \citep{2021ApJS..254...22J},
python-FSPS \citep{2014zndo.....12157F}, 
SEDpy \citep{2021zndo...4582723J}, 
matplotlib \citep{2007CSE.....9...90H}, 
seaborn \citep{2021JOSS....6.3021W}, 
jupyter/IPython Notebooks \citep{2016ppap.book...87K}, 
SAOImage DS9 \citep{2003ASPC..295..489J}, 
pyRAF \citep{2012ascl.soft07011S}, 
Source Extractor \citep{1996A&AS..117..393B}}






\appendix

\begin{table}[htbp]
  \centering
  \caption{Free parameters used in \texttt{Prospector} SED fitting}
    \begin{tabular}{p{0.1\linewidth}p{0.49\linewidth}p{0.41\linewidth}}
    \toprule
    Parameter            & Description          & Priors \\
    \midrule
    $z_{\rm obs}$        & Observed redshift (initalized to mean $z$ from \citet{2020ApJS..247...25B}) & TopHat: [$z-0.001$, $z+0.001$] \\
    M$_{\rm BCG}$ (M$_{\odot}$) & Total stellar mass formed & Log$_{10}$ uniform: [$10^9, 10^{13}$] \\
    $\log(Z/Z_{\odot})$  & Stellar metallicity in log solar units & Clipped normal: $\mu = 0.0$, $\sigma = 0.3$, range=[$-2.0, 0.5$] \\
    $t_{\rm age}$        & Age of galaxy        & TopHat: [$0$, age of universe at $z_{\rm obs}$] \\
    $\tau$               & e-folding time of star formation history in Gyr & Log$_{10}$ uniform: [$0.01, 3.0$] \\
    $d_2$                & Optical depth for stellar light attenuation by dust for old stars using extinction curve from \citet[][eqn. 1]{2013ApJ...775L..16K}, where observed flux $I = I_0 e^{-d_2}$ & TopHat: [$0, 20$] \\
    dust ratio           & Ratio to convert from dust attenuation optical depth from old stars ($d_2$) to additional attenuation for young stars & Clipped normal: $\mu = 2.3$, $\sigma = 0.3$, range=[$0, 3.0$] \\
    $\delta$             & Dust index in \citet{2013ApJ...775L..16K} extinction curve (eqn. 1)  & TopHat: [$-2.0, 0.5$] \\
    $f_{\rm burst}$      & Fraction of total stellar mass formed in a recent star formation burst & TopHat: [$0, 0.5$] \\
    $f_{\rm age, burst}$ & Time at which burst happens, as a fraction of $t_{\rm age}$ & TopHat: [$0.1, 1$] \\
    $\sigma_v$           & Velocity smoothing in km s$^{-1}$ & TopHat: [$150, 500$] \\
    spec$_{\rm norm}$    & Spectrum normalization factor to match photometry & Log$_{10}$ uniform: [$0.005, 5.0$] \\
    spec$_{\rm jitter}$  & Part of pixel outlier mixture model, to marginalize over poorly modeled noise like residual sky lines or missing absorption lines & TopHat: [$1,10$] \\
    spec$_{\rm outlier}$ & Multiplicative noise inflation term & TopHat: [$0.0001,1$] \\
    ($p_1$, $p_2$, $p_3$) & Continuum calibration (Chebyshev) polynomial & TopHat: n=3: [$-0.2/(n+1), 0.2/(n+1)$] \\
    \bottomrule
    \end{tabular}%
  \label{tab:priors}%
\end{table}%


\listofchanges

\end{document}